\documentclass{article}

\usepackage{spconf}
\usepackage{graphicx}
\usepackage[utf8]{inputenc} % allow utf-8 input
\usepackage[T1]{fontenc}    % use 8-bit T1 fonts
\usepackage{hyperref}       % hyperlinks
\usepackage{url}            % simple URL typesetting
\usepackage{booktabs}       % professional-quality tables
\usepackage{amsfonts}       % blackboard math symbols
\usepackage{nicefrac}       % compact symbols for 1/2, etc.
\usepackage{microtype}      % microtypography
\usepackage[dvipsnames]{xcolor}         % colors
\usepackage{todonotes}
\usepackage{tabularx}
\usepackage{float}
\usepackage[square,sort,comma,numbers]{natbib}
\usepackage{setspace}

\definecolor{colora}{RGB}{216,27,96}
\definecolor{colorb}{RGB}{30,136,229}
\definecolor{colorc}{RGB}{255,193,7}
\definecolor{colord}{RGB}{0,77,64}

\title{Multi-Modal Pre-Training for Automated Speech Recognition}

\name{David M. Chan$^{\star \dagger 1}$ \qquad Shalini Ghosh$^{\dagger}$ \qquad Debmalya Chakrabarty$^{\dagger}$ \qquad Bj{\"o}rn Hoffmeister$^{\dagger}$}
\address{$^{\star}$ University of California, Berkeley \\
     	 $^{\dagger}$ Amazon Alexa AI}
      
\begin{document}
%\ninept
%
\maketitle

\begin{abstract}
	 Traditionally, research in automated speech recognition has focused on local-first encoding of audio representations to predict the spoken phonemes in an utterance. Unfortunately, approaches relying on such hyper-local information tend to be vulnerable to both local-level corruption (such as audio-frame drops, or loud noises) and global-level noise (such as environmental noise, or background noise) that has not been seen during training. In this work, we introduce a novel approach that leverages a self-supervised learning technique based on masked language modeling to compute a global, multi-modal encoding of the environment in which the utterance occurs. We then use a new deep-fusion framework to integrate this global context into a traditional ASR method, and demonstrate that the resulting method can outperform baseline methods by up to 7\% on Librispeech; gains on internal datasets range from 6\% (on larger models) to 45\% (on smaller models).
%  Traditionally, methods for automated speech recognition focus on local-first representations by extracting phonemes directly from small (~10ms) segments of audio. In a lab setting with clean audio, such representations can be sufficient to achieve good performance, however, in real-world automated speech recognition, small aberrations in the audio or extreme background noise that has not been experienced during training can lead to poor generalization performance for models. In this work, we explore several methods to introduce global-level audio representations
\end{abstract}

\section{Introduction}

Despite considerable research, automated speech recognition (ASR) remains an extremely challenging task, especially in noisy environments. Correctly understanding spoken phonemes requires an understanding of speech patterns, as well as an understanding of myriad varieties of background noise, much of which may never have been encountered by a model during the training process. Many traditional ASR methods such as the RNN-T and Conformer \citep{graves2013speech, zhang2020transformer, gulati2020conformer} focus on a local understanding of phonemes predicted from small 10-30ms segments on audio. Unfortunately, such local-first representations may leave ASR models vulnerable to extreme noise such as frame drops or sudden loud noises at the local level, as demonstrated by Chiu et al. \citep{chiu2021rnn}. Not only are such models vulnerable to local disruptions, they can be affected as well by global-level noise unseen during training.

In this paper, we target the problem of such global-level noise in utterances. Many ASR datasets such as Librispeech \cite{panayotov2015librispeech} are collected in lab-specific environments with a canonical set of situations, which leaves a long tailed distribution of noisy environments uncovered. While local level disruptions can in part be solved by introducing semantic-level language modeling \cite{gulati2020conformer, alayrac2020self}, the global out-of-distribution noise problem has no such simple solution. While self-supervised learning has been used in the student/teacher context in ASR \cite{watanabe2017student, zhang2020semi, manohar2018teacher}, we speculate (and believe that it is important future work to confirm) that the additional performance gained by student teachers is a direct response to exposure to a large tail of environmental effects, regardless of the ASR content. 

\begin{figure}
\centering
	\includegraphics[width=0.85\linewidth]{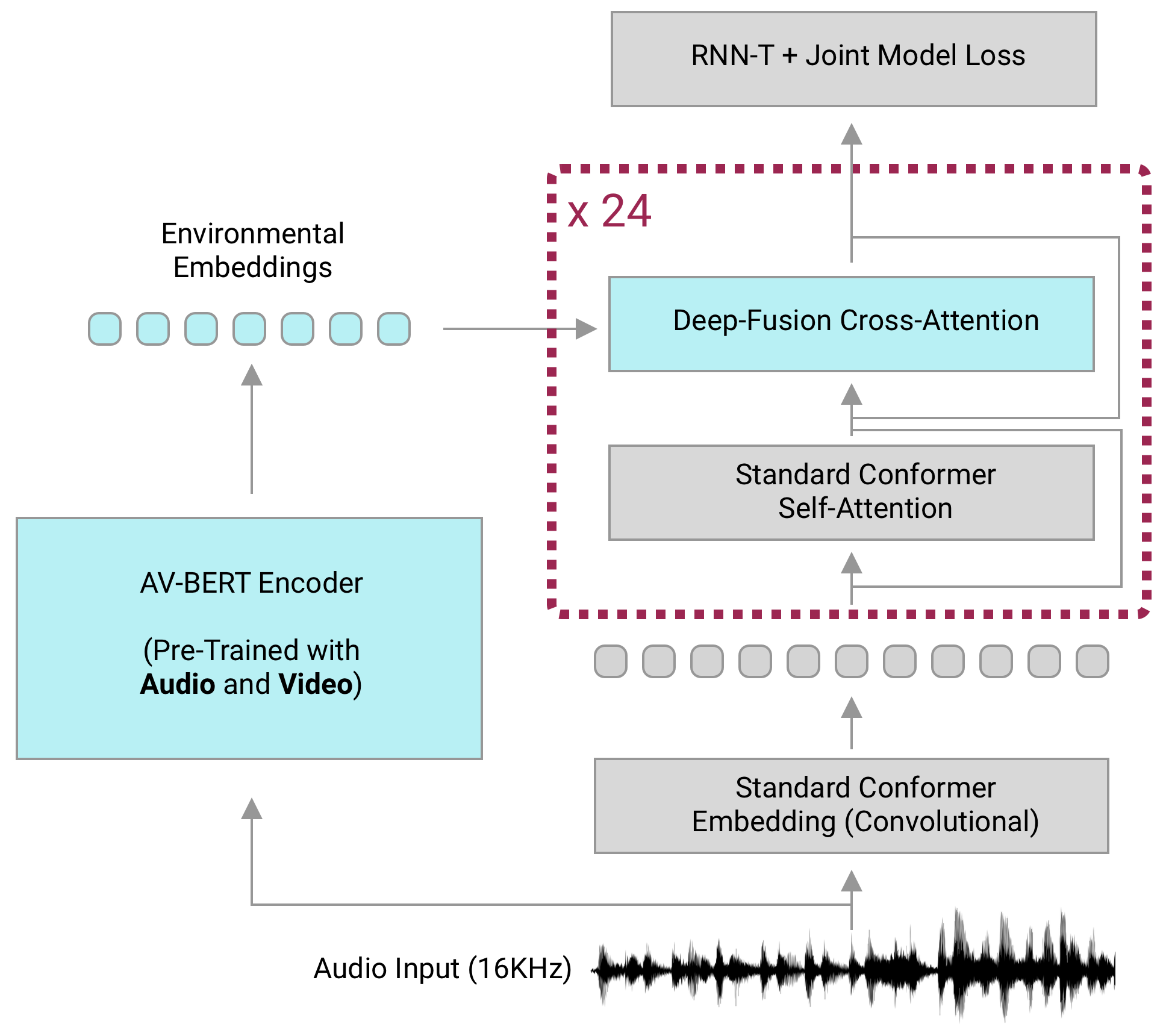}
	\caption{An overview of our proposed approach to the ASR training process using deep-fusion with environmental embeddings. Our audio is fed to the pre-trained environmental representation model, trained on large-scale multimodal data. We then use a stack of 24 deep-fusion cross-attention layers in the base conformer architecture to deeply fuse the environmental representations with a standard conformer model. The RNN-T and joint model loss remain unchanged from \cite{gulati2020conformer}.}
	\label{fig:econformer}
	\vspace{-1.2em}
\end{figure}

Recently, both vision and NLP communities have introduced several methods \cite{akbari2021vatt, devlin2018bert, sun2019videobert} which make use of large unlabeled data to build self-supervised representations transferable to downstream tasks. Such representations can both provide exposure to the long-tailed data distribution and reduce the required labeled training data in transfer \cite{sun2019videobert, hsu2021hubert}. Notably, Hsu et. al. \cite{hsu2021hubert} recently introduced HuBERT, which shows that general representations of audio are beneficial for ASR tasks.

Our proposed method goes beyond HuBERT, and expands the context of environment-level global representation to both audio \textbf{and} visual data. We hypothesize that such visual data provides pseudo-labels, which help to weakly annotate auditory features, allowing the model to capture robust information in the long-tail. Recently \cite{wang2021multimodal} confirmed that even in the absence of the video representation at test time, audio-video self-supervised model representations can outperform audio-only models on \textit{audio only} tasks. Concurrently with this work, \cite{shi2022learning} modified HuBERT for multi-modal downstream tasks but did not focus on the potential applications to ASR. 
 
In designing our proposed method, we exploit the ability of visual information to organize audio representations, as demonstrated by Wang et. al. \cite{wang2021multimodal}. However, unlike the model proposed by Wang et al., our method leverages masked language modeling (MLM) as a joint training objective, as opposed to contrastive representation learning. In contrastive representation learning, samples from the audio and video domains are pushed into a joint latent space at a global level. This mode of training, however, \textit{inherently suggests that the modalities should lie in the same latent space}, which we believe to be sub-optimal for automated speech recognition, where we want to focus on globally aware local-first representations (one phoneme should retain its' independence of other phonemes, but the representation should still be contextually aware). On the other hand, MLM, popularized in BERT \citep{devlin2018bert} and further extended to multiple modalities in VideoBERT \citep{sun2019videobert} and UniT \citep{hu2021unit}, focuses on contextualizing local representations with the ability to reconstruct the full sequence, leading to local-first global aware representations.

In an effort to address the global representation problem, we (1) introduce a multi-modal pre-training scheme (AV-BERT) based on masked language modeling (Section \ref{sec:avbert}), (2) develop a novel deep-fusion scheme for training on joint local/global representations in the ASR domain (Section \ref{sec:econformer}) and (3) demonstrate the benefits of generating robust environmental representations with visual information, even when no visual information is present at test time (Section \ref{sec:results}).

\section{Approach}
\label{sec:methods}

%\begin{figure*}[h]
%	\includegraphics[width=\linewidth]{sections/figures/overview.png}
%	\caption{An overview of our approach. In the first stage, we train a joint audio/video representation learner using masked language modeling and K-Means quantization over the raw video and audio inputs (Section \ref{sec:avbert}, Figure \ref{fig:avbert}). In the second stage, we augment a standard conformer encoder with the learned representations using a deep cross-attentional fusion approach (See Section \ref{sec:econformer}, Figure \ref{fig:econformer}). Using a two-stage approach allows us to train strong, locally-contextual environment-level representations in the first stage, and then leverage these representations \textit{as necessary} in the second stage.}
%	\label{fig:overview}
%\end{figure*}

Our method consists of a two-stage approach (an overview is given in Figure \ref{fig:econformer}) inspired by ideas from \cite{sun2019videobert}, \cite{hsu2021hubert} and \cite{alayrac2020self}. In the first stage we build a video-augmented audio representation using a pre-training task based on masked language modeling. In the second stage, we use these video-augmented audio representations to provide additional context to a conformer-based ASR model. 

\subsection{Multimodal Pre-Training}
\label{sec:exp_det}
\label{sec:avbertpproc}
\label{sec:avbert}

\begin{figure*}
	\centering
	\includegraphics[width=\linewidth]{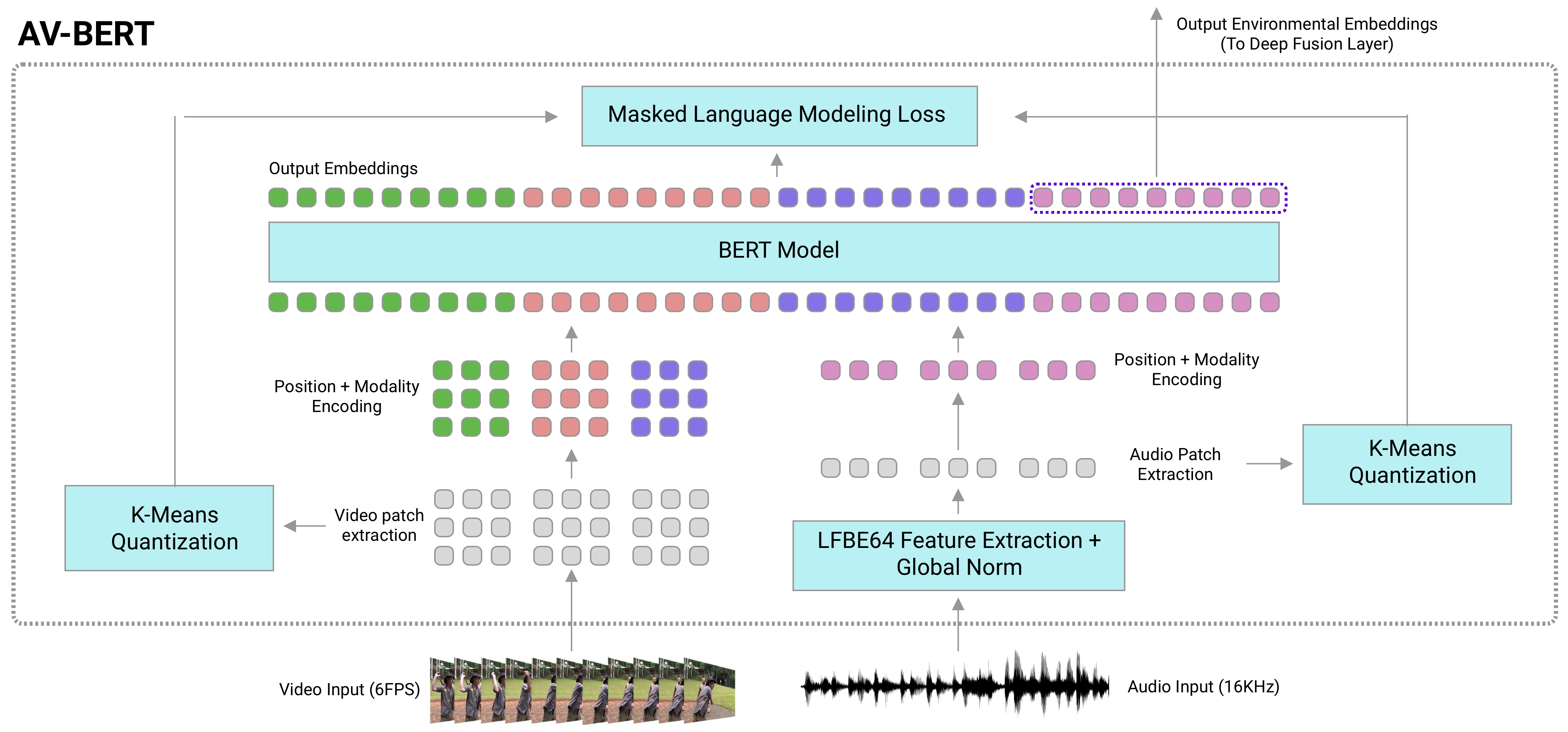}
	\caption{An overview of our pre-training model. First, a set of patches are extracted from the multimodal inputs. Next, these patches are quantized using k-means and embedded directly using convolutional layers, modality encodings, and positional encodings. The embedded patches form the input sequence, which is passed to a standard BERT masked-language model. The quantized token labels are used along with the output of the masked BERT model to perform masked-language prediction.}
	\label{fig:avbert}
	\vspace{-1em}
\end{figure*}

While many methods for learning multimodal representations focus on self-supervised learning with a contrastive objective, our proposed method AV-BERT differs in that it uses a masked language modeling objective. Our pre-training encoder model, shown in Figure \ref{fig:avbert}, takes inspiration from the UniT \cite{hu2021unit} model for unified transformer architectures, however instead of using multiple encoders, we use a single unified encoder layer. We take further inspiration from ViViT \cite{arnab2021vivit} and use a video-patch based encoding to the transformer, while taking inspiration from HuBERT's \cite{manohar2018teacher} iterated quantization training method for masked-language modeling from raw signals. In this section, we dive deeper into each of the components, and discuss our modeling choices from the perspective of an ASR-first multimodal pre-training model.

To build a multimodal representation learning method based on masked language modeling principles, we first consider a token-based representation of modalities. We draw the representation from a discrete quantization of both the video and audio domains. 

%For the video modality, we extract non-overlapping patches of the input data of shape $(3, 16, 16, a)$ (3 frames, 16x16 patches across the RGB channels). Each of these patches are then quantized using a mini-batch K-Means clustering method with 8192 clusters, trained on the Kinetics training set for 20 epochs using a batch size of 512. For the audio representation, we extract non-overlapping patches of shape $(3, 192)$ from our initial data with shape $(Frames, 192)$. These patches are also quantized using mini-batch K-Means with 4096 clusters. We combine both quantizations into a single unified language with 12288 tokens.

For the video modality, we extract non-overlapping voxels of shape $3 \times 16 \times 16$ of the input data, and 3-frame patches of audio data. These patches are then quantized using a K-means model with $8192$ video centers, and $4096$ audio centers trained offline. While we could use the quantized tokens directly as input to the masked language model as was done in VideoBERT \cite{sun2019videobert}, similar to HuBERT \cite{hsu2021hubert} we use the raw data as input while classifying based on the quantization. This allows the model to see the full input audio/video, and also allows the model to respond to subtle changes in the input which cannot be captured by our audio-visual language (which only consists of a total of 12288 audio/video tokens).

Thus, as input to our model, we use a set of modality-specific convolutions (matching the patch dimensions) to embed both the video and audio in the model dimension. We then apply a learned modality embedding, as well as learned position embedding \cite{devlin2018bert}. For the audio, we use the frame index as the position. For the video, we apply a spatio-temporal position embedding identical to that from Timesformer \cite{bertasius2021space}. We then flatten the spatial dimensions, and concatenate the video and audio sequences along the temporal axis, to form the input to the multimodal BERT-style encoder.

To perform masked language modeling, we use an architecture similar to BERT \cite{devlin2018bert}, which allows for full cross-attention between all points (both in the audio and video modalities, as well as spatiotemporally). This can lead to very long sequence lengths, so we compensate by reducing the per-node batch size and distributing the model across several GPUs. Because of the distribution across many GPUs, we do not use batch normalization --- we instead use instance normalization to enhance training efficiency, since distributed batch normalization requires cross-GPU communication and is under-defined for small per-node batch sizes.

The training of the AV-BERT model is heavily dependent on the choice of masking technique. If we mask tokens uniformly with some rate, it is unlikely that the model will learn cross-model representations, as both audio and video are highly local representations (information in one location tends to share a large amount of mutual information with other neighbors). A naive approach would be to mask entire modalities at a time, however often the modalities are not heavily correlated enough to reconstruct a quantized representation of the other. Instead, we apply a progressive masking technique along the sequence dimension, where we begin the training by masking local information to encourage local-level representations, and progressively increase the size of the masks during training. This encourages the model to first learn local representations, and then eventually learn more global representations as the training process continues. Explicitly, we initialize the masking with a random probability of $0.15$ and a mask width of 1. As training progresses, we increase the mask width and probability, ending with a final mask width of 11 and a mask probability at any center of $0.45$. The probability is increased on a exponential decay schedule over 10,000 steps --- every time we reach 10,000 optimization steps, the mask width is increased and the masking probability is reset.

We perform our pre-training using the publicly available splits of the Kinetics-600 dataset \cite{carreira2018short}. The Kinetics-600 dataset consists of 366K 10 second videos, each with a corresponding audio track and an associated action label (from 600 classes). For video, we reduce the frame-rate to 6FPS and resize the short side of the video to 256 pixels, and take a 256x256 center crop of the resulting resized video. For the audio, we resample the raw input audio to 16KHz, and stack 3 adjacent Log-Filterbank Energy features of dimension 64 (LFBE-64) from the resulting audio frames. The features are whitened using the mean and standard deviation computed globally from the training data, then clipped to lie between $(-1.2,1.2)$ for numerical stability.

In our pre-training experiments, we use a model dimension of 128 with six BERT encoder blocks. The model is implemented in Tensorflow and is trained using 32 Nvidia-V100 GPUs, each with a batch size of 8, for 100 epochs (or until the validation perplexity converges, whichever comes first). To perform the optimization, we use an Adam \cite{kingma2014adam} optimizer with a fixed learning rate of $3e^{-4}$, and $\beta_1=0.9,\beta_2=0.99$.

\subsection{Automated Speech Recognition Downstream Task}
\label{sec:econformer}
\label{sec:asrdata}

For the downstream automated speech recognition task, we have two goals: (a) maintain the performance of current state of the art machine learning techniques, and (b) augment the current models with additional global-level environment context to improve phoneme recognition. In order to accomplish these goals, we modify the conformer architecture (as shown in Figure \ref{fig:econformer}) to include additional cross-attention layers, which attend across the vector-level representations generated by our pre-trained AV-BERT model. This method allows the model to selectively pay attention to global context information learned by our model, while preserving the local-first approach favored by ASR techniques. This helps resolve one of the major challenges faced by HuBERT \cite{hsu2021hubert} in the ASR domain: when you focus on learning global representations, you can fail to encode the information necessary for local-first tasks such as phoneme detection. During the training of the downstream model, we freeze the representations learned by AV-BERT, to both reduce the computational complexity and to maintain a more global representation level even after significant training. Note that AV-BERT does not introduce additional trainable parameters, as AV-BERT is trained offline and only the audio representations are leveraged during the ASR training.

We evaluate the proposed model on the LibriSpeech \cite{panayotov2015librispeech} dataset, which consists of 970 hours of labeled speech. Because our audio embedding method is frozen during the training process, to ensure that there is no domain shift we follow the same audio pre-processing technique as in Section \ref{sec:avbertpproc} with additional SpecAugment \cite{park2019specaugment}. In addition to Librispeech, we present results on several internal datasets: ``Base" representing general speech, ``Query", representing standard speech queries, ``Rare" representing the long-tailed distribution of rare words, and ``Messages" representing longer message-based utterances. All customer-specific data has been de-identified from these internal datasets. For these results, we use the same model architecture; however once pre-training of AV-BERT is complete, we fine-tune our final ASR model on a corpus consisting of 120K hours of labeled speech and 180K hours of unsupervised speech, using in-house teacher distillation.

For the ASR Conformer, we use a hidden dimension of 1024 and 24 self-attention/cross-attention blocks, with convolutional downsampling on the input waveform with a kernel size of 3 and a stride of 2. We use an identical joint model to Gulati et al. \cite{gulati2020conformer} with a graph-based ASR decoding framework. The optimization hyper-parameters are shared with AV-BERT and described in Section \ref{sec:exp_det}, with the exception that we use a per-node batch size of 28.

\section{Results}
\label{sec:results}

\begin{figure}
    \footnotesize
    \setlength\tabcolsep{0.5pt} % default value: 6pt
    \centering
    \begin{tabularx}{\linewidth}{lXl}
	    \toprule
	    \textbf{Reference:}&& should  i  buy  from  the  princess  starfrost  set     royale  high\\
	    Base (M):&& should  i  buy  from  the  princess  \textcolor{colorc}{stare}  \textcolor{colorc}{froset}  \textcolor{colora}{in}  \textcolor{colora}{we're}  \textcolor{colora}{all}  \textcolor{colora}{rawhide}\\
	    Ours (M):&& should  i  buy  from  the  princess  \textcolor{colora}{star} \textcolor{colorc}{frost}  set     royale  high\\
	    %$\quad$ A/V + Conformer (L) && add & to & o & c & playlist \\
    \end{tabularx}
    \begin{tabularx}{\linewidth}{lXl}
	    \toprule
	    \textbf{Reference:}&& read  all  of  lisa  left  eye  lopes  songs  including  the  thirteen  more\\
	    Base (M):&& read  all  of  lisa  \textcolor{colorb}{****}  \textcolor{colorb}{***}  \textcolor{colora}{loeb}   songs  including  the  thirteen  \textcolor{colora}{horn}\\
	    Ours (M):&& read  all  of  lisa  left  eye  \textcolor{colora}{lopez}  songs  including  the  thirteen  more\\
	    %$\quad$ A/V + Conformer (L) && add & to & o & c & playlist \\
    \end{tabularx} 
    \begin{tabularx}{\linewidth}{lXl}
	    \toprule
	    \textbf{Reference:}&& ... signalman he lead tenor for telephone wires so soldiers ...\\
	    Base (M):&& ... \textcolor{colora}{signal map he'd late tenoff} telephone \textcolor{colora}{wise} \textcolor{colorb}{**} soldiers ...\\
	    Ours (M):&& ... signalman he lead \textcolor{colora}{teno} for telephone wires so soldiers...\\
	    %$\quad$ A/V + Conformer (L) && add & to & o & c & playlist \\
	    \bottomrule
    \end{tabularx} 
    \caption{Examples showing improvements on long utterances in our model, vs the baseline model. \textcolor{colorb}{Blue} indicates deletions, \textcolor{colora}{pink} indicates substitutions and \textcolor{colorc}{yellow} indicates insertions.}
    \label{fig:qualitative}
    \vspace{-1em}
\end{figure}

Our main results are presented in Table \ref{tab:results}. We report results on two models, a model using AV-BERT trained with both the video and audio components of the Kinetics dataset, as well as a model trained with only the audio from Kinetics. In Table \ref{tab:results}, the Baseline model is identical to the proposed model except all multi-modal cross-attention layers are replaced with self-attention layers to preserve the parameter count. 

The results show that both models — one trained on audio only and another with audio + video embeddings — outperform the baseline model. We further validate the method with experiments on internal Alexa AI datasets in Table \ref{tab:results-alexa}.Our fine-tuned model is warm-started from the base Alexa model (300K hours of audio), and augmented with the additional training on the 377K samples. The benefit is more pronounced in smaller models, as the contextual representations are more useful with fewer training data, and parameters.

\begin{table}
\footnotesize
\caption{\small Results summary of word error rate for the Librispeech dataset with no additional language model. The baseline model replaces the cross-attentions with self-attention (to closely preserve parameters) using the same training profile (See Section \ref{sec:methods}). ``A" is the audio-only model, and ``A/V" is the full Audio/Video BERT.}\label{tab:results}
\vspace{0.7em}
\begin{tabularx}{\linewidth}{llll}
	\toprule
		\textbf{Method} & \textbf{Params (M)} & \textbf{test-clean} & \textbf{test-other} \\
	\midrule
	\textbf{LAS} & & &  \\
		$\quad$ Transformer	\cite{synnaeve2019end}				& 370		& 2.89	& 	6.98 \\
		$\quad$ Transformer \cite{karita2019comparative}			& -			& 2.2	& 	5.6  \\
		$\quad$ LSTM \cite{gulati2020conformer}					& 360		& 2.6 	& 	6.0  \\
	\textbf{Transducer} & & &  \\
		$\quad$ Transformer 	\cite{zhang2020transformer}			& 139		& 2.4	& 	5.6  \\
		%$\quad$ ContextNet (S) \cite{han2020contextnet}			& 10.8		& 2.9	& 	7.0  \\
		$\quad$ ContextNet (M) \cite{han2020contextnet} 			& 31.4		& 2.4	& 	5.4  \\
		$\quad$ ContextNet (L) \cite{han2020contextnet}			& 112.7 & 2.1	& 	4.6  \\
	\textbf{Conformer} & & &  \\
		%$\quad$ Conformer (S) \cite{gulati2020conformer}  		& 10.3		& 2.7	& 	6.3 \\
		$\quad$ Conformer (M) \cite{gulati2020conformer}  		& 30.7		& 2.3	& 	5.0 \\
		$\quad$ Conformer (L) \cite{gulati2020conformer} 		& 118.8		& 2.1	& 	4.3 \\
	\midrule
	\textbf{Ours} & & & \\
		$\quad$ Conf. (M, base)	& 79 		& 2.21	& 	4.85 \\
		$\quad$ Conf. (L, base)	& 122 		& 2.11	& 	4.29 \\
		$\quad$ A + Conf. (M)		& 79			& 2.15 \textbf{\textcolor{colord}{\footnotesize{(+2.7\%)}}}		& 	4.82 \textbf{\textcolor{colord}{\footnotesize{(+0.6\%)}}} \\
		$\quad$ A/V + Conf. (M)		& 79			& 2.10 \textbf{\textcolor{colord}{\footnotesize{(+4.8\%)}}} 	& 	4.72 \textbf{\textcolor{colord}{\footnotesize{(+2.7\%)}}} \\
		$\quad$ A/V + Conf. (L)		& 122		& 1.98	\textbf{\textcolor{colord}{\footnotesize{(+7.0\%)}}} & 	4.10 \textbf{\textcolor{colord}{\footnotesize{(+4.4\%)}}} \\
	\bottomrule
\end{tabularx}
\vspace{-1.5em}
\end{table}

\begin{table}
\footnotesize
\caption{\small Relative improvement over baseline WER for Alexa-AI datasets methods without a language model.}\label{tab:results-alexa}
\vspace{0.7em}
\begin{tabularx}{\linewidth}{lrrrr}
	\toprule
		\textbf{Method} & \textbf{Base} & \textbf{Rare} & \textbf{Query} & \textbf{Messages} \\
	\midrule
	\textbf{Conformer (M)}			& 0 		& 0		& 	0 & 0 \\
		$\quad$ + Audio (M)				& +30.1\%		& +17.9\%		&  	+26.7\% & +20.1\% \\
		$\quad$ + Audio/Video (M)		& +45.6\%		& +31.2\%		& 	+38.7\% & +17.2\%\\
	\textbf{Conformer (L)}			& 0 		& 0		& 	0  & 0\\
		%$\quad$ + Audio (L)		& -		& -		& 	- \\
		$\quad$ + Audio/Video (L)		& +5.1\%		& +5.4\%		& 	+4.2\% & +5.9\%\\
	\bottomrule
\end{tabularx}
\vspace{-1em}
\end{table}

\section{Conclusion}
\label{sec:conclusion}

In this paper, we have introduced an initial approach for exploring global-level contextual embeddings for the automated speech recognition pipeline. We build a novel self-supervised vision + audio encoder, and demonstrate the performance of this method by using  deep-fusion to directly connect the contextual embeddings with the local-first conformer model. Our model demonstrates strong performance on Librispeech, and presents a new direction for exploration into multimodal ASR.

\clearpage
\section{References}
\begingroup
  \def\section*#1{}
  \small
  \setlength{\bibsep}{4pt}
  \bibliographystyle{IEEE}
  \bibliography{references}

\begin{thebibliography}{10}

\bibitem{graves2013speech}
Alex Graves, Abdel-rahman Mohamed, and Geoffrey Hinton,
\newblock ``Speech recognition with deep recurrent neural networks,''
\newblock in {\em 2013 IEEE international conference on acoustics, speech and
  signal processing}. Ieee, 2013, pp. 6645--6649.

\bibitem{zhang2020transformer}
Qian Zhang, Han Lu, Hasim Sak, Anshuman Tripathi, Erik McDermott, Stephen Koo,
  and Shankar Kumar,
\newblock ``Transformer transducer: A streamable speech recognition model with
  transformer encoders and rnn-t loss,''
\newblock in {\em ICASSP 2020-2020 IEEE International Conference on Acoustics,
  Speech and Signal Processing (ICASSP)}. IEEE, 2020, pp. 7829--7833.

\bibitem{gulati2020conformer}
Anmol Gulati, James Qin, Chung-Cheng Chiu, Niki Parmar, Yu~Zhang, Jiahui Yu,
  Wei Han, Shibo Wang, Zhengdong Zhang, Yonghui Wu, et~al.,
\newblock ``Conformer: Convolution-augmented transformer for speech
  recognition,''
\newblock {\em arXiv preprint arXiv:2005.08100}, 2020.

\bibitem{chiu2021rnn}
Chung-Cheng Chiu, Arun Narayanan, Wei Han, Rohit Prabhavalkar, Yu~Zhang,
  Navdeep Jaitly, Ruoming Pang, Tara~N Sainath, Patrick Nguyen, Liangliang Cao,
  et~al.,
\newblock ``Rnn-t models fail to generalize to out-of-domain audio: Causes and
  solutions,''
\newblock in {\em 2021 IEEE Spoken Language Technology Workshop (SLT)}. IEEE,
  2021, pp. 873--880.

\bibitem{panayotov2015librispeech}
Vassil Panayotov, Guoguo Chen, Daniel Povey, and Sanjeev Khudanpur,
\newblock ``Librispeech: an asr corpus based on public domain audio books,''
\newblock in {\em 2015 IEEE international conference on acoustics, speech and
  signal processing (ICASSP)}. IEEE, 2015, pp. 5206--5210.

\bibitem{alayrac2020self}
Jean-Baptiste Alayrac, Adria Recasens, Rosalia Schneider, Relja Arandjelovic,
  Jason Ramapuram, Jeffrey De~Fauw, Lucas Smaira, Sander Dieleman, and Andrew
  Zisserman,
\newblock ``Self-supervised multimodal versatile networks.,''
\newblock {\em NeurIPS}, vol. 2, no. 6, pp. 7, 2020.

\bibitem{watanabe2017student}
Shinji Watanabe, Takaaki Hori, Jonathan Le~Roux, and John~R Hershey,
\newblock ``Student-teacher network learning with enhanced features,''
\newblock in {\em 2017 IEEE International Conference on Acoustics, Speech and
  Signal Processing (ICASSP)}. IEEE, 2017, pp. 5275--5279.

\bibitem{zhang2020semi}
Zi-qiang Zhang, Yan Song, Jian-shu Zhang, Ian~Vince McLoughlin, and Li-rong
  Dai,
\newblock ``Semi-supervised end-to-end asr via teacher-student learning with
  conditional posterior distribution.,''
\newblock in {\em INTERSPEECH}, 2020, pp. 3580--3584.

\bibitem{manohar2018teacher}
Vimal Manohar, Pegah Ghahremani, Daniel Povey, and Sanjeev Khudanpur,
\newblock ``A teacher-student learning approach for unsupervised domain
  adaptation of sequence-trained asr models,''
\newblock in {\em 2018 IEEE Spoken Language Technology Workshop (SLT)}. IEEE,
  2018, pp. 250--257.

\bibitem{akbari2021vatt}
Hassan Akbari, Linagzhe Yuan, Rui Qian, Wei-Hong Chuang, Shih-Fu Chang, Yin
  Cui, and Boqing Gong,
\newblock ``Vatt: Transformers for multimodal self-supervised learning from raw
  video, audio and text,''
\newblock {\em arXiv preprint arXiv:2104.11178}, 2021.

\bibitem{devlin2018bert}
Jacob Devlin, Ming-Wei Chang, Kenton Lee, and Kristina Toutanova,
\newblock ``Bert: Pre-training of deep bidirectional transformers for language
  understanding,''
\newblock {\em arXiv preprint arXiv:1810.04805}, 2018.

\bibitem{sun2019videobert}
Chen Sun, Austin Myers, Carl Vondrick, Kevin Murphy, and Cordelia Schmid,
\newblock ``Videobert: A joint model for video and language representation
  learning,''
\newblock in {\em Proceedings of the IEEE/CVF International Conference on
  Computer Vision}, 2019, pp. 7464--7473.

\bibitem{hsu2021hubert}
Wei-Ning Hsu, Yao-Hung~Hubert Tsai, Benjamin Bolte, Ruslan Salakhutdinov, and
  Abdelrahman Mohamed,
\newblock ``Hubert: How much can a bad teacher benefit asr pre-training?,''
\newblock in {\em ICASSP 2021-2021 IEEE International Conference on Acoustics,
  Speech and Signal Processing (ICASSP)}. IEEE, 2021, pp. 6533--6537.

\bibitem{wang2021multimodal}
Luyu Wang, Pauline Luc, Adria Recasens, Jean-Baptiste Alayrac, and Aaron
  van~den Oord,
\newblock ``Multimodal self-supervised learning of general audio
  representations,''
\newblock {\em arXiv preprint arXiv:2104.12807}, 2021.

\bibitem{shi2022learning}
Bowen Shi, Wei-Ning Hsu, Kushal Lakhotia, and Abdelrahman Mohamed,
\newblock ``Learning audio-visual speech representation by masked multimodal
  cluster prediction,''
\newblock {\em arXiv preprint arXiv:2201.02184}, 2022.

\bibitem{hu2021unit}
Ronghang Hu and Amanpreet Singh,
\newblock ``Unit: Multimodal multitask learning with a unified transformer,''
\newblock {\em arXiv preprint arXiv:2102.10772}, 2021.

\bibitem{arnab2021vivit}
Anurag Arnab, Mostafa Dehghani, Georg Heigold, Chen Sun, Mario Lu{\v{c}}i{\'c},
  and Cordelia Schmid,
\newblock ``Vivit: A video vision transformer,''
\newblock {\em arXiv preprint arXiv:2103.15691}, 2021.

\bibitem{bertasius2021space}
Gedas Bertasius, Heng Wang, and Lorenzo Torresani,
\newblock ``Is space-time attention all you need for video understanding?,''
\newblock {\em arXiv preprint arXiv:2102.05095}, 2021.

\bibitem{carreira2018short}
Joao Carreira, Eric Noland, Andras Banki-Horvath, Chloe Hillier, and Andrew
  Zisserman,
\newblock ``A short note about kinetics-600,''
\newblock {\em arXiv preprint arXiv:1808.01340}, 2018.

\bibitem{kingma2014adam}
Diederik~P Kingma and Jimmy Ba,
\newblock ``Adam: A method for stochastic optimization,''
\newblock {\em arXiv preprint arXiv:1412.6980}, 2014.

\bibitem{park2019specaugment}
Daniel~S Park, William Chan, Yu~Zhang, Chung-Cheng Chiu, Barret Zoph, Ekin~D
  Cubuk, and Quoc~V Le,
\newblock ``Specaugment: A simple data augmentation method for automatic speech
  recognition,''
\newblock {\em arXiv preprint arXiv:1904.08779}, 2019.

\bibitem{synnaeve2019end}
Gabriel Synnaeve, Qiantong Xu, Jacob Kahn, Tatiana Likhomanenko, Edouard Grave,
  Vineel Pratap, Anuroop Sriram, Vitaliy Liptchinsky, and Ronan Collobert,
\newblock ``End-to-end asr: from supervised to semi-supervised learning with
  modern architectures,''
\newblock {\em arXiv preprint arXiv:1911.08460}, 2019.

\bibitem{karita2019comparative}
Shigeki Karita, Nanxin Chen, Tomoki Hayashi, Takaaki Hori, Hirofumi Inaguma,
  Ziyan Jiang, Masao Someki, Nelson Enrique~Yalta Soplin, Ryuichi Yamamoto,
  Xiaofei Wang, et~al.,
\newblock ``A comparative study on transformer vs rnn in speech applications,''
\newblock in {\em 2019 IEEE Automatic Speech Recognition and Understanding
  Workshop (ASRU)}. IEEE, 2019, pp. 449--456.

\bibitem{han2020contextnet}
Wei Han, Zhengdong Zhang, Yu~Zhang, Jiahui Yu, Chung-Cheng Chiu, James Qin,
  Anmol Gulati, Ruoming Pang, and Yonghui Wu,
\newblock ``Contextnet: Improving convolutional neural networks for automatic
  speech recognition with global context,''
\newblock {\em arXiv preprint arXiv:2005.03191}, 2020.

\end{thebibliography}
\endgroup

\end{document}